%% file: main_paper.tex
\documentclass[sigconf]{acmart}

\AtBeginDocument{%
  \providecommand\BibTeX{{%
    \normalfont B\kern-0.5em{\scshape i\kern-0.25em b}\kern-0.8em\TeX}}}

\copyrightyear{2023}
\acmYear{2023}
\setcopyright{rightsretained}
\acmConference[SIGCSE 2023]{Proceedings of the 54th ACM Technical Symposium on Computer Science Education V. 1}{March 15--18, 2023}{Toronto, ON, Canada} \acmBooktitle{Proceedings of the 54th ACM Technical Symposium on Computer Science Education V. 1 (SIGCSE 2023), March 15--18, 2023, Toronto, ON, Canada} 
\acmDOI{10.1145/3545945.3569885} 
\acmISBN{978-1-4503-9431-4/23/03}

\newcommand{\displaynothing}[1]{}
\usepackage{enumitem}
\usepackage{graphicx}
\setkeys{Gin}{width=\linewidth,totalheight=\textheight,keepaspectratio}
\graphicspath{{graphics/}}
\usepackage{tikz}
\usepackage{algorithm}
\usepackage{hyperref}
\usepackage[noend]{algpseudocode} 
\usepackage{multirow}
\definecolor{darkblue}{rgb}{0, 0.5, 0.9}
\definecolor{darkgreen}{rgb}{0.1, 0.7, 0.4}

\usepackage[leftmargin=3em]{quoting}

\definecolor{todocolor}{rgb}{0.8,0,0}

\makeatletter
\gdef\@copyrightpermission{
 \begin{minipage}{0.3\columnwidth}
  \href{https://creativecommons.org/licenses/by/4.0/}{\includegraphics[width=0.90\textwidth]{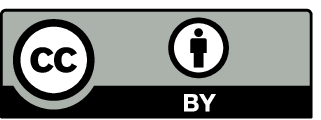}}
 \end{minipage}\hfill
 \begin{minipage}{0.7\columnwidth}
  \href{https://creativecommons.org/licenses/by/4.0/}{This work is licensed under a Creative Commons Attribution International 4.0 License.}
 \end{minipage}
 \vspace{5pt}
}

\newcommand\footnoteref[1]{\protected@xdef\@thefnmark{\ref{#1}}\@footnotemark}
\makeatother

\begin{document}

\title{Inclusive Study Group Formation At Scale}


\author{Sumer Kohli}
\email{sumer.kohli@berkeley.edu}
\authornotemark[1]
\affiliation{%
  \institution{University of California, Berkeley}
  \country{}
}

\author{Neelesh Ramachandran}
\email{neelesh.r@berkeley.edu}
\authornotemark[1]
\affiliation{%
  \institution{University of California, Berkeley}
  \country{}
}

\author{Ana Tudor}
\email{anamtudor@berkeley.edu}
\authornotemark[1]
\affiliation{%
  \institution{University of California, Berkeley}
  \country{}
}

\author{Gloria Tumushabe}
\email{gloriatumushabe@berkeley.edu}
\affiliation{%
  \institution{University of California, Berkeley}
  \country{}
}

\authornote{These authors contributed equally to the research.}

\author{Olivia Hsu}
\email{owhsu@stanford.edu}
\affiliation{%
  \institution{Stanford University}
  \country{}
}

\author{Gireeja Ranade}
\email{ranade@eecs.berkeley.edu}
\affiliation{%
  \institution{University of California, Berkeley}
  \country{}
}









\renewcommand{\shortauthors}{Sumer Kohli et al.}


\begin{abstract}

Underrepresented students face many significant challenges in their education. In particular, they often have a harder time than their peers from majority groups in building long-term high-quality study groups. This challenge is exacerbated in remote-learning scenarios, where students are unable to meet face-to-face and must rely on pre-existing networks for social support. 

We present a scalable system that removes structural obstacles faced by underrepresented students and supports all students in building inclusive and flexible study groups. One of our main goals is to make the traditionally informal and unstructured process of finding study groups for homework more equitable by providing a uniform but lightweight structure. We aim to provide students from underrepresented groups an experience that is similar in quality to that of students from majority groups. Our process is unique in that it allows students the opportunity to request group reassignments during the semester if they wish. Unlike other collaboration tools our system is not mandatory and does not use peer-evaluation.

We trialed our approach in a 1000+ student introductory Engineering and Computer Science course that was conducted entirely online during the COVID-19 pandemic. We find that students from underrepresented backgrounds were more likely to ask for group-matching support compared to students from majority groups. At the same time, underrepresented students that we matched into study groups had group experiences that were comparable to students we matched from majority groups. B-range students in high-comfort and high-quality groups had improved learning outcomes.

\end{abstract}


\maketitle

\input{sections/introduction}

\input{sections/methods}
\input{sections/analysis}

\vspace{-2mm}
\begin{acks}
Thank you to all UC Berkeley EECS 16A Fall 2020 students and staff for making this possible, as well as to Armando Fox, Tesha Sengupta-Irving, Thomas Philip, Aleata Hubbard Cheuoua, Austin Patel, Drew Kaul, and Rahul Arya for their help. Thank you to Google for their support through a 2022 Award for Inclusion Research Grant.
\end{acks}

\bibliographystyle{ACM-Reference-Format}
\bibliography{final_ref}

\clearpage
\appendix

\end{document}

%% file: sections/introduction.tex
\section{Introduction}

 
In 2013, a news outlet~\cite{eastbayexpress} ran a story about the University of California, Berkeley, 
with a quote from the then director of the African American Student Development office:
\begin{quoting}[leftmargin=2mm]
\textit{``A black student might be in a science course, ... and the professor says, `Okay, everybody has to have a study group.' Nobody picks them for a study group. They first have to show that they can get an A before they get selected.''}
\end{quoting}
%
This problem still persists. In 2020 the president of the Black engineering students association at UC Berkeley~\cite{carlos} wrote: 
\begin{quoting}[leftmargin=2mm]
\textit{``... I've stood by black CS undergrads who were constantly ostracized in group projects, often the last ones to be picked. Their peers regularly perceive them to be less knowledgeable and capable in carrying coding projects. I have had to console crying teenagers scarred from the experience, questioning, ..., whether being black was worth it. Imagine that. Their only path to recognition in that environment was through a bogus burden of excellence. What ever happened to the opportunity of being average yet respected?''}
\end{quoting}

These quotes illustrate some of the challenges faced by underrepresented students on college campuses. The social isolation of underrepresented students is a ``systemic problem''~\cite{hafferty2011can}. Additionally, underrepresented students can face an undue burden of excellence~\cite{hodge2008theorizing}, a burden of representing their demographic group~\cite{thompson2002being}, or stereotype threat~\cite{steele2011whistling}. In group settings, students may experience bias and microaggressions~\cite{grindstaff2019no, ong2011inside},  \emph{solo-status}~\cite{thompson2002being, spangler1978token} (\emph{i.e.}, be the only member of their demographic group present in an otherwise homogeneous group) or \emph{tokenism}~\cite{spangler1978token}. 

Peer groups~\cite{turetsky2020psychological,  mishra2020social} and social networks~\cite{calvo2004effects, thomas2000ties} can have a large effect on the success of individuals. While some students come to college with a pre-established network, many students from underrepresented groups may not know other students when they start college. This, combined with the other factors mentioned, can make it much harder for a student from an underrepresented group to initiate a study group as compared to a student from a majority group. Despite the value of these study groups, many students simply do not have them. 


\begin{table*}[t]
\begin{center}
\begin{tabular}{ |c|c| } 
 \hline
\textbf{Challenge} & \textbf{Approach to address challenge} \\ 
\hline
Study group initiation burden & Lightweight process, integrated into course structure \\ \hline
Bias & Uniform process for all students   \\ \hline
Solo-status & Algorithmically prevented \\ \hline
Microaggressions & Structured guidelines, instructor-initiated groups encourage professionalism
\\ 
 & Reassignment process offers agency and recourse in case of negative experiences\\
 \hline
 Stereotype threat &  No peer-evaluation or connection to grades \\
 \hline
\end{tabular}
\end{center}
\caption{Summary table indicating our approach to various challenges faced by students}
\label{tab:one}
\vspace{-8mm}
\end{table*}



We hypothesize that by providing a standardized structure for study group formation, we can mitigate some of the exclusion faced by minoritized students. Our goal is to help students in building long-term (one semester or more) study groups that evolve into friendships and support systems. For this, we design a process that is (1) \emph{scalable}, in that it can be used for classes with thousands of students with minimal instructor intervention, (2) \emph{inclusive}, in that it attempts to remove hurdles faced by underrepresented students in forming study groups, and (3) \emph{flexible}, in that it is opt-in and provides students the agency to request reassignment as desired. 
%
%
%

\vspace{-4mm}
\subsection{Previous work on group formation}
The fields of computer-supported collaborative learning (CSCL) and computer science education have explored various avenues for using technology to support student collaboration~\cite{wise2017visions, uttamchandani2020finding}. However, critics of CSCL point out that the tangible impact of the field has been small~\cite{wise2017visions}, and some~\cite{uttamchandani2020finding} urge the field to \emph{center on equity} to promote research that can make a difference. 

Forming good student groups is a complex multi-objective combinatorial optimization problem, where one must balance student schedules, personalities, skills, learning outcomes, instructor preferences, fairness and more. As a result, there has been tremendous work on different strategies and algorithms for group formation. Some recent surveys give an overview of much of the work around student teams~\cite{cruz2014group, odo2019group,putro2018group,borges2017group,maqtary2019group}. 
%
For examples, some group formation strategies are based on topic-mastery and skills~\cite{deibel2005team, clarke1994pieces, dzvonyar2018team, mujkanovic2016improving, catme, hertz_2022}, others on problem-solving or learning styles~\cite{adan2011forming}, schedules~\cite{wills1994peer, dzvonyar2018team, catme, hertz_2022}, personality and interpersonal skills~\cite{lambic2018novel,mujkanovic2016improving}, and demographics~\cite{catme,hertz_2022, lovold2020forming}. Additionally, people have considered both algorithmic and machine-learning approaches to team formation~\cite{dzvonyar2018algorithmically, mujkanovic2012unsupervised, lambic2018novel}. Team assignments have been considered in the contexts of both final projects for MOOCs~\cite{staubitz2019graded, ju2018scalable} and short-term activities~\cite{srba2014dynamic}. Some of these works have also provided tools that can be used by others~\cite{catme, hertz_2022}.

At this point, we would like to make a distinction between \emph{study-groups} (that are informal, fluid, open-ended, and student-driven, e.g. meeting with friends in your dorm to work), from \emph{project-groups} (that are officially tied to grades, mandatory, inflexible, operate for a specific time, and have a predetermined focus). The prior work, including all of the works mentioned above, has largely focused on instructor-mandated groups that are assessment-focused, and are more like project groups than study-groups. In addition to considering group composition, they often focus on building teamwork skills in students. For instance, the most widely-used tool (8000+ instructors) is likely the CATME Teammaker~\cite{catme}. This team-formation software offers a built-in framework for peer-evaluation~\cite{aseepeer33497}. While such peer-evaluation can be very helpful for developing teamwork skills~\cite{ju2018scalable,oakley2004turning}, and can even improve grades~\cite{oneill2020team}, it can be difficult to implement effectively~\cite{van2010effective}, and can have negative implications for student relationships~\cite{zou2018student}. In the context of research-peer-reviews, unprofessional reviews may disproportionately harm underrepresented groups~\cite{silbiger2019unprofessional}. 
The focus on developing teamwork skills, and constantly being evaluated, can trigger stereotype threat and detract from the social support that a more informal study group would provide~\cite{mosley2007stigma}. 

\emph{This work:} Our opt-in system builds non-mandatory (a student may choose to form their own group, ask us to form a group, or have no group at all) study groups, as opposed to project groups, with careful considerations for the experience of underrepresented students. A novelty of our system is that it offers students the chance to be reassigned to a new group after trying out a group for some time. We avoid groups where students may experience solo-status~\cite{thompson2002being}, and also consider intersectional identities~\cite{ Solomon2018NotJB, Rankin2020}. We also try to avoid homogeneous groups, given the importance of diversity in teams~\cite{hong2004groups}, and the value of increased social networks~\cite{calvo2004effects}. 
To the best of our knowledge, such optional (but instructor-generated) study groups that are not related to student assessment have not been studied extensively in the literature.

\vspace{-3mm}
\subsection{Our contributions}
We design a lightweight system for study group formation that is integrated into the course structure and thus standardized for all students. We hypothesize that this reduces the \emph{structural barriers} faced by underrepresented students in creating study groups, while also supporting \emph{all} students. By creating instructor-supported but student-controlled opportunities for discussing coursework, we create opportunities for repeated interaction and proximity~\cite{rohrer2021proximity}, which can lead to friendship~\cite{preciado2012does}, while setting social norms that encourage positive interactions~\cite{koppenhaver2003structuring, dixon2009establishing}. The voluntary participation and ability to request reassignments gives students agency that can be essential in learning~\cite{stenalt2021does}. Table~\ref{tab:one} summarizes our approaches.

Large classes can be particularly advantageous for forming study groups, because small demographic groups can be amplified. In a class of 1000 students, if 30\% are women and 1\% are Black (approximate percentages at UC Berkeley), that is 300 women and 10 Black students. This scale allows for a flexibility in group formation that is not possible in a 100 student class with the same proportions. 

We test our approach in a 1000+ student introductory undergraduate class in Engineering and Computer Science that was conducted remotely due to the COVID-19 pandemic\footnote{IRB approval for Protocol ID `2020-08-13526' and Approval Date `September 25, 2020.'}. The large course size allows us to compute statistically significant results in many cases, which is not possible in some of the smaller studies discussed earlier. Given the demographics in our classroom (typical for our institution), in this analysis we consider students identifying as women or gender non-conforming, as well as students identifying as Black, African American, Hispanic, Native American, Alaskan Native, Hawaiian Native (see numbers in Fig.~\ref{fig:self_form_comp}) as underrepresented. To support adoption of our system, all of the processes, surveys and code used for this process are available open-source~\cite{our_repo}, unlike systems such as CATME~\cite{catme}. 

\emph{Questions.} To understand if our system created positive experiences for underrepresented students, we ask the following questions.
(1) Who participates in the structured group formation process? Who participates in the reassignment process? (2)
Does the structured group formation process lead to students from minoritized and majority groups having similar experiences in their group interactions? (3)
Does the structured group formation process impact student performance on exams?

%% file: sections/methods.tex
\section{Methods}

\subsection{High-level system description}
Our system design was informed by informal informational interviews with cultural and identity-based STEM student organizations
and the Director of Student Diversity in our department. Given their informal nature, we do not report on the interviews here. Students mentioned that one of the biggest challenges is trying to find people to work with when many other students already have connections with each other. It was also mentioned that it can be difficult for women to find women-majority groups to work in, that scheduling times for study-groups is hard, and microaggressions in study groups can lead to people wanting to work alone. The Director of Student Diversity emphasized the longer-term impacts of study-groups such as influencing post-graduation decisions as well as increasing enjoyment in education.

We focused on a process that was simple and low-overhead. There were no grade incentives associated with joining groups. The process was opt-in, and the students had the choice of forming their own groups (we call these \emph{self-matched} groups), asking us to form a group for them (we call these \emph{software-matched} groups), or having no group at all.

\paragraph{Timeline} Students opted-in to be matched to a group through a survey on their first homework assignment in Week 1. Reassignment rounds were conducted in Weeks 5 and 9, where students were invited to request new groups by updating their survey preferences if they were not happy with the first group they were matched to. This was also an opportunity for students to provide feedback on the matching process/ group experience thus far. A final evaluation survey to collect student feedback on their experience was released in Week 13, through the final homework.

\paragraph{Group-creation survey} \label{sec:survey-design}
Groups were created based on students' scheduling preferences and other information, collected in Week 1. In particular, the student survey collected information about (1) current timezone (our students were distributed around the globe due to COVID-19), (2) preferred meeting days, (3) preferred meeting times on those days, (4) year (\emph{e.g.,} freshman, sophomore etc.), (5) other courses taken in the current semester (to maximize overlap), (6) whether student previously took a linear algebra or a coding class, (7) preferred discussion section time (so students could attend the same discussion section), (8) how important the course is to the student, and  (9) gender and race identity. Questions (7), (8) and (9) were optional. Students were encouraged to select-all-that-apply for scheduling questions to maximize matching potential, so someone that is free on both Monday and Tuesday would select both options. The exact surveys for group-creation and evaluation, and the group guidelines discussed below are available at~\cite{our_repo}. 

\paragraph{Supporting in-group interactions} Following work such as~\cite{oakley2004turning}, we released a set of guidelines for collaboration~\cite{our_repo}. This included rotating role-assignments (Facilitator, Timekeeper, Librarian, Scribe) and suggestions for the meeting structure. 
Additionally, we designed new open-ended homework questions to encourage discussion and sharing of ideas as coursework-related ``ice-breakers.''

\subsection{Evaluation survey} \label{sec:evalsurvey}
We evaluated the student experience using a survey~\cite{our_repo} in Week 13. 
Five questions related to the quality of the study group experience for each student: (1) the comfort of the student in asking questions in their group, (2) the comfort of the student in sharing ideas with their group, (3) the interaction frequency (how many times a week did they interact), (4) the number of students in the study group that regularly participated, and (5) whether the student wants to take future courses with their group. We also collected information about the number of reassignments requested by students and the obstacles they experienced in the success of their study groups. 

\subsection{Group formation algorithm} \label{sec:algorithm}
As discussed earlier, group formation is a multi-objective optimization problem with fairness and resource constraints. There are no clear optimal strategies for this problem, and multiple approaches such as greedily maximizing heuristics ~\cite{catme,odo2019group},  genetic algorithms~\cite{hertz_2022}, clustering approaches~\cite{odo2019group}, similarity maximization~\cite{odo2019group}, and more have been tried. Here, we choose to take a partitioning-based approach with an explicit priority-ordering of the constraints, instead of an indirect prioritization through an optimization heuristic. 

At a high-level, our algorithm repeatedly partitions the class based on scheduling and course-matching constraints, and then ensures that group-composition demographic constraints are met in the formed groups. 
We formed groups of four to six students~\cite{oakley2004turning}, with some exceptions to maximize matching flexibility and avoid solo-status. 
The algorithm is divided into three phases: (1) Multi-partitioning, (2) Tentative group assignment, and (3) Finalizing group composition to avoid solo-status. 

\label{subsec:multi-partitioning}
\textbf{Multi-partitioning:} 
\input{figs_code/partition_tree_example}
Multi-partitioning repeatedly partitions the class until all matching criteria are exhausted. For Fall 2020, we used the following priority-ordered partitioning sequence: (1) timezone, (2) meeting day, (3) meeting time, (4) class year, (5) courses this semester, (6) previous courses, (7) discussion time, and (8) course importance. If some of the partitions are too small (less than four students), they are merged into the nearest neighboring partitions (distance computed as edge hops on the partition tree). 

Student schedules are the most important constraint. To maximize matching flexibility, we allow a student to be provisionally placed into \emph{all} the partitions that their schedule allows. For example, someone who is free on Monday morning, Monday afternoon and Tuesday afternoon would be provisionally placed in all three partitions. Fig.~\ref{fig:multi-partition_tree} shows an example of the tree formed by this process, using student timezone as the first partitioning criterion and then meeting day as the second criterion. Duplicates are handled after tentative group assignment: once a student is assigned to a group, they are removed from all other partitions.



\textbf{Tentative group assignment:}
We iterate over the partitions to assign groups. We first assign groups in the smallest partitions that are ``just right,'' \textit{i.e.,} those with four to six students in them. Once a group is formed, the corresponding students are removed from all other partitions. If the group assignment causes a partition to dip below the smallest viable group size, the partition is merged into the nearest neighboring partition. Partitions are considered in increasing order of size after that. In partitions with multiple groupings possible, groups are formed randomly. 

\textbf{Finalizing group-composition to avoid solo-status:} 
The last step in the algorithm considers the racial and gender identities of the students in the groups. In particular, it is at this stage that we ensure that no students have solo-status (\textit{i.e.,} are the only member of their race or gender) in their groups. To achieve this, the algorithm iterates over all groups to identify any students with solo-status. Students with solo-status are removed from their tentative assignments and 
partnered into pairs or triples of students who share the same gender and race identities, while considering intersectional identities. If suitable partners cannot be found within the partition, we allow partnering outside the partition. Once the pairs/triples in each partition are generated, they are randomly merged with another pair/triple to create a study group. We first search within the partition for a pair/triple to merge with, but if none exist, we extend our search to other neighboring partitions. 
Since the course composition may be such that it might not be possible to avoid solo-status for \emph{all} students (\textit{e.g.,} if there is only one student from a particular racial group), this process is stopped after the number of students with solo-status is below a configurable threshold.
Finally, we manually checked to ensure that there were no racially-homogeneous groups~\cite{hong2004groups} (some gender homogeneity is unavoidable), which turned out to be the case. However, this is not always guaranteed. 
Our code is available at~\cite{our_repo}. 

%% file: figs_code/partition_tree_example.tex
\begin{figure}
\begin{center}
    \begin{tikzpicture}[->, level distance=1.4cm, sibling distance = 1.4cm, thick, scale=0.8, transform shape]
     
        \node[draw, circle, fill=darkblue!25]{All}  
        child {node [draw, circle, fill=red!25]{Pacific} 
            child { node [draw, circle, fill=darkgreen!50]{Mo} }
            child { node [draw, circle, fill=darkgreen!50]{Tu} }
            child { node [draw, circle, fill=darkgreen!50]{We} }
            child { node {...} }
            child [missing]
            child [missing]
        }
        child {node [draw, circle, fill=red!25]{Eastern}
            child [missing]
            child { node [draw, circle, fill=darkgreen!50]{Mo} }
            child { node {...} }
        }
        child {node [draw, circle, fill=red!25]{Europe}
            child [missing]
            child [missing]
            child { node[draw, circle, fill=darkgreen!50] {Mo} }
            child { node {...} }
        }
        child {node [draw, circle, fill=red!25]{$\ldots$}
            child [missing]
            child [missing]
            child { node[draw, circle, fill=darkgreen!50] {$\ldots$} }
        };
        
        \node[draw=none, darkblue] at (-4, 0) {\textbf{[All Students]}};
        \draw[->, darkblue] (-2.8, 0) -- (-0.8,0);
        \node[draw=none, red!70] at (-4, -0.7) {\textbf{[Timezone Partitions]}};
        \draw[->, red!70] (-4, -0.9) -- (-2.8,-1.2);
        \node[draw=none, darkgreen] at (-5.2, -1.6) {\textbf{[Scheduling Partitions]}};
        \draw[->, darkgreen] (-5, -1.8) -- (-5,-2.2);
    \end{tikzpicture}
    \caption{\label{fig:multi-partition_tree}An example of class partitions modeled in a tree-like figure. If a student \textit{could} be part of multiple partitions, they are initially placed into \textit{all of these partitions}, which are later pruned to ensure that each student is part of one group. }
\end{center}
\vspace{-5mm}
\end{figure}
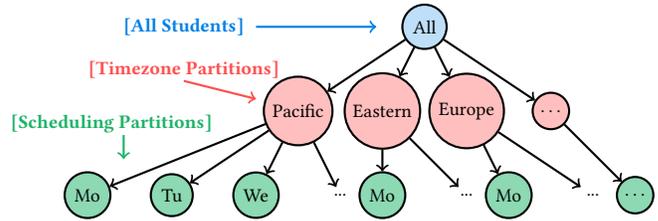

%% file: sections/analysis.tex
\vspace{-4mm}
\section{Analysis and Evaluation}
\label{sec:evaluation}
\newcommand{\placeholderfig}[1]{{\color{red} PLACEHOLDER FIG TITLE: #1}}

Our analysis is based on responses to the student preference survey, reassignment requests, the final evaluation survey, and grade information. We only consider the 477 students who consented to participate in this research, though many more students participated in the group matching process. 

We consider each of the five questions on the evaluation survey (Sec.~\ref{sec:evalsurvey}) as a group-quality indicator, and we classify positive responses as below:
\begin{enumerate}
\item Comfort asking questions: Students agreed or strongly agreed that they feel comfortable asking questions in the group
    \item Comfort sharing ideas: Students agreed or strongly agreed that they feel comfortable sharing ideas in the group
    \item Interaction frequency: students interacted with their group at least once a week 
    \item Participation: Some, most, or all members participated
      \item Future courses: students state they hope they can, or definitely will, take future courses with their group
    
\end{enumerate}

\subsection{Participation demographics} \label{sec:partdemo}
\input{figs_code/table_demographic_distribution}
Fig.~\ref{fig:self_form_comp} showcases different demographic breakdowns of our participants, as well as breakdowns across those who chose self-matched vs. software-matched groups. 
143 students reported having self-matched groups and 334 students asked for software-matched groups. 
Looking at the percentages in of students who requested software-matched groups in Column (C) in Fig.~\ref{fig:self_form_comp}, we see that women and gender-nonconforming/ genderqueer students were more likely to request software-matched study groups than men, and Black, Hispanic and White students were more likely to request software-matched groups than Asian students (the largest demographic group).

\emph{Requesting reassignments.} 
27\% of students requested one reassignment and 2\% of students requested two reassignments. 
Significantly fewer Asian students requested reassignment (24\%) compared to non-Asian and non-White students (39\%). Hispanic students (48\%) (and White students (41\%)) requested reassignments significantly more often than non-Hispanic (28\%) (non-White (26\%)) students, respectively, indicating that initial group matches did not work out as well for these students. Significance in all these cases was verified using 2-sample proportion z-tests, $\alpha = 0.05$~\cite{rice2006mathematical}.

\emph{Software-matched students.}
Software-matched students had overall positive group experiences. For example, 74\% and 78\% of students reported feeling comfortable asking questions and sharing ideas respectively. We performed significance tests in comparison to the null hypothesis that 50\% of students would have positive study group experiences and 50\% of students would have negative study group experiences. Using a 2-sample proportion z-test ($\alpha=0.05$)~\cite{rice2006mathematical}, over 50\% of all software-matched students had positive experiences with group interaction, group activity, and comfort asking questions and sharing ideas. 

We note that positive responses to the evaluation questions tended to correlate with each other, using the Pearson correlation coefficient. Specifically, positive responses to the group interaction and group participation questions were correlated with each other ($r = 0.83$) . Additionally, the comfort indicators correlated with each other ($r = 0.79$).

\emph{Self-matched students.} We find that self-matched students had even more positive experiences than software-matched students (93\% and 95\% report comfort in asking questions and sharing ideas respectively), similar to the findings of~\cite{lovold2020forming}. This is not surprising since many of the self-matched groups are formed by students who knew each other before college. The remaining analysis focuses on software-matched students.


\subsection{Impact on underrepresented students}
In general, students in software-matched groups who identify as being from an underrepresented racial or gender demographic did not demonstrate significant differences in responses compared to majority demographic groups, with a few exceptions.  

\begin{figure}[h]
    \centering
    \includegraphics{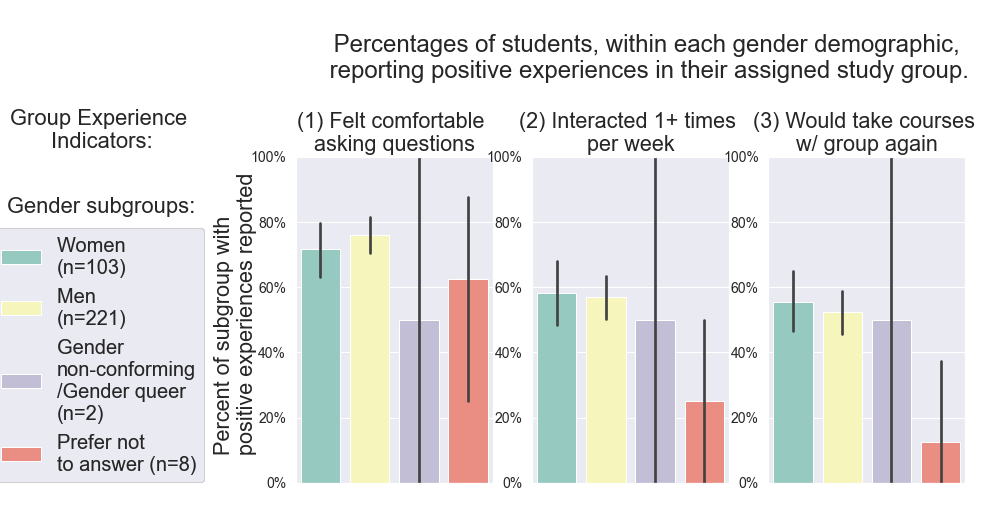}
    \caption{Responses by gender for three representative questions from the evaluation survey (software-matched students). The percentage of positive responses is on the $Y$-axis, the black bar is an error bar.}
    \label{fig:gender_boxplots}
    \vspace{-3mm}
\end{figure}

\subsubsection{Gender}
Figure~\ref{fig:gender_boxplots} shows responses to representative survey questions for different gender groups. Each vertical bar corresponds to the percentage of positive responses to the question for a particular gender. Bars that are roughly the same height indicate similar experiences across gender groups. 

Using 2-sample proportion z-tests ($\alpha = 0.05$)~\cite{rice2006mathematical}, we found no significant difference between students in the majority group (men) and those who identified as any other gender (Fig.~\ref{fig:gender_boxplots}). 

One student identifying as gender non-conforming / genderqueer (GNC) had positive experiences in all categories, and one other GNC-identifying student had negative experiences in all categories, but the small sample size  prevents any generalizable conclusions.

\begin{figure}[h]
    \centering
    \includegraphics{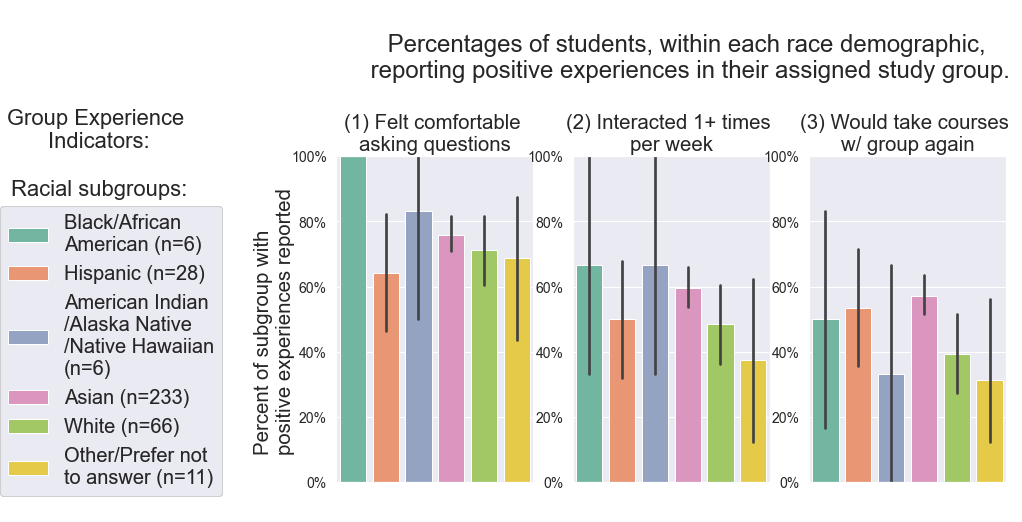}
    \caption{Responses by race for the same questions for software-matched students. 
    Black/AA (green bar) and Hispanic (orange bar) students respond similarly to other groups. }
    \label{fig:race_boxplots}
    \vspace{-2mm}
\end{figure}

\subsubsection{Race}
In this analysis students who self-identified with multiple racial groups are counted in each of the categories.
Fig.~\ref{fig:race_boxplots} shows responses to representative survey questions for different racial groups, in a similar style to Fig.~\ref{fig:gender_boxplots}, for software-matched students. 
We see that students identifying as Black / African-American (AA) did not show statistically significant differences  in any study group indicators, in comparison to non-Black/AA students (using a Fisher exact test, $\alpha =0.05$, since the sample sizes are small~\cite{rice2006mathematical}). All six Black/AA students in the class reported positive indicators.

Similarly, Hispanic\footnote{We recognize that Hispanic is an ethnicity, and not a racial group, per the US Census Bureau, however we include it here to accurately reflect our survey.}-identifying students did not show statistically significant differences from non-Hispanic students (Fisher exact test, $\alpha =0.05$~\cite{rice2006mathematical}). One interesting observation is that even though White students reported similar comfort levels to non-White students, they were significantly less likely than others to want to take courses with their groups again (2-sample proportion z-test, $\alpha=0.05$~\cite{rice2006mathematical}) (Fig.~\ref{fig:race_boxplots}, subfigure 3), which requires further investigation, especially given that White and Hispanic students requested reassignments significantly more often (see Sec.~\ref{sec:partdemo} ). We were unable to make general conclusions about the experiences of students identifying as Native American, Alaska Native, or Native Hawaiian due to the small sample size and mixed responses.

\subsubsection{Student year}
\begin{figure}[h]
    \centering
    \includegraphics{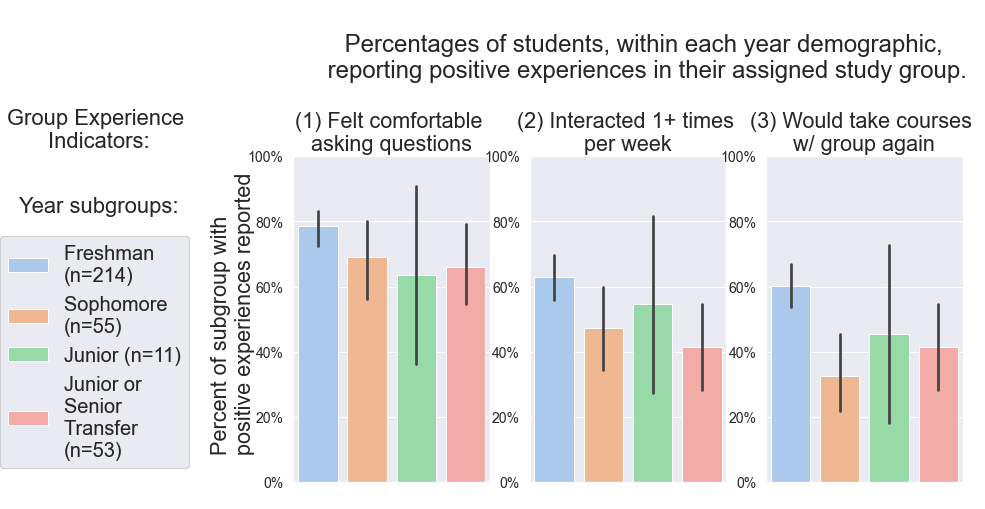}
    \caption{Responses by year for the three representative questions for software-matched students. Freshmen students respond significantly more positively for all questions.} 
    \label{fig:year_boxplots}
    \vspace{-3mm}
\end{figure}

Freshmen students had overwhelmingly positive study group experiences (Fig.~\ref{fig:year_boxplots}), with significantly higher proportions of positive responses to all study group indicators in comparison to non-freshmen (2-sample proportion z-test,  $\alpha=0.05$~\cite{rice2006mathematical}). This may because freshmen were more enthusiastic overall, and the course is targeted to freshmen.

\subsection{Exam performance}
\label{section:grades_assoc}

We find that just participation in a study group, whether self-matched or software-matched, did not correlate with higher exam scores for students. It makes sense that the specific experiences in the group matter. We consider a student to be in a \emph{high-comfort} study groups if they responded positively to at least one of indicators 1 or 2, and \emph{high-activity} study groups as those where students response positively to at least one of indicators 3 or 4~(using indicators defined at the beginning of Sec.~\ref{sec:evaluation}). 

We identify several correlations with software-matched high-comfort and high-activity groups. Students in high-comfort groups had higher final exam scores (average = 72.2\%) than others (average = 66.1\%) (p = 0.011, Student's t-test~\cite{rice2006mathematical}). We examined more closely student performance in different grade ranges. 
 Students with B-range scores (68-89\%) on the first midterm (we call these B-range students) who were also in high-comfort groups saw significantly higher final exam grades than B-range students not in high-comfort groups (72.2 \% vs. 63.9\%, p = 0.004, Student's t-test~\cite{rice2006mathematical}). Similar gains were also observed on the final exam for B-range students in high-activity groups vs. those not in high-activity groups (71.7\% vs. 65.8\%, p=0.008, Student's t-test~\cite{rice2006mathematical}). Students who received A, and C to below C-level, grades on MT1 had no significant associations between their scores and group quality indicators. 
 
 
 
\subsection{Study-group obstacles}
34.5\% of students reported no major obstacles in their study groups. Despite the fact that our matching system explicitly prioritized student availability, students struggled to schedule group meetings, and 48.5\% of students reported this as a challenge they faced. Over 31.7\% reported that Zoom fatigue obstructed the success of their meetings and 30.7\% also found interactions with students they did not know awkward. Of students listing other reasons, multiple students indicated that lack of responsiveness from their group mates was a challenge in establishing the study group. 
 

%


\section{Limitations}
Unfortunately, we are limited in generalized conclusions due to the small sample sizes of some of the demographic groups, which is the reality of many CS courses. Informal anecdotal feedback from students in underrepresented groups was positive, and we plan to conduct formal focus groups in the future. 
Our setting makes it difficult to run a randomized control study, since it would be unethical to disallow students from joining a study group. 
The experiment was conducted during a remote semester, and further work will be required to explore how these conclusions hold in the context of in-person instruction. The PI for the study was also the instructor for the course, and strongly encouraged group participation, which may have increased participation. 
There are multiple sub-optimalities in the implementation of our partitioning algorithm. In particular, the process of merging small partitions together could be optimized based on partition size. Partnering to avoid solo-status could be done before group formation as well. We can also optimize the random assignment in large partitions. 

\vspace{-2mm}
\section{Discussion}
This work provides an example of how a structured approach to study group formation can provide similar experiences for students from different demographic groups. This system is clearly filling an unmet need for underrepresented students, given the differential rates at which Black, Hispanic and women students requested study groups as compared to their peers. 

For those students who did participate in successful study groups, experiences were similar across demographic groups. High-comfort and high-activity groups were further associated with improved learning outcomes. While these findings might be explained by external factors, e.g., a student’s academic proactiveness and general social comfort may be associated with their better performance on exams, the positive associations are promising. The large scale of the experiment allowed us to have statistically significant conclusions in many (but not all) cases. 

Student scheduling and non-responsive students are two of the biggest obstacles to successful study groups. The majority of reassignment requests came from students whose initial assigned group members were non-responsive. We are improving our system to automatically send reminder emails to improve this issue. In some cases, there was a lack of leadership, where no one initiated meetings, which lead to the group not working out. We are considering explicitly asking students to self-identify as leaders in the initial preferences survey to address this. Many students also reported that they wished they had reached out to other students more, so active encouragement of this through guidelines and lecture might also be valuable. This all points to the need to have more active support during the semester for keeping the study groups active. While our current work does not focus on developing team dynamics, we hope to bring in best practices to encourage successful collaborations such as those in~\cite{oakley2004turning}.

The challenges around scheduling were surprising to us, since these were explicitly prioritized by the matching algorithm. We learned anecdotally that changing schedules and preferences on part of the students might be an issue here. This suggests an intervention where we might want to try slotting in ``study group time'' into official course schedules so that students who want to be in a group can block off that time when they sign up for courses.

%% file: figs_code/table_demographic_distribution.tex
\begin{figure}
    \centering
    \begin{tabular}{|l|llll|}
        \hline
        \textbf{Demographic group} & \textbf{(A)} & \textbf{(B)} & \textbf{(C)} & \textbf{(D)}              \\ \hline \hline
        Women             & 139 & 103 & 74.1\%  &(66.8\%, 81.4\%)      \\ \hline
        Men               & 326 & 221 & 67.8\%  &(62.7\%, 72.9\%)      \\ \hline
        \begin{tabular}[c]{@{}l@{}}Gender non-conforming/\\ Genderqueer \end{tabular} & 2   & 2   & 100\% & - \\ \hline
        Other/Prefer not to answer & 10   & 10   & 100\% & - \\ \hline \hline
        \begin{tabular}[c]{@{}l@{}}Black/ African American\end{tabular}                           & 7  & 6  & 85.7\% & (59.8\%, 100\%)  \\ \hline
        Hispanic          & 39  & 28  & 71.8\% & (57.7\%, 85.9\%) \\ \hline
        \begin{tabular}[c]{@{}l@{}}Native American/ \\Alaska Native/\\ Hawaiian Native \end{tabular} & 9  & 6  & 66.7\% & (35.9\%, 97.5\%) \\ \hline
        White             & 86  & 66  & 76.7\% & (67.8\%, 85.7\%) \\ \hline
        Asian             & 345 & 233 & 67.5\% & (62.6\%, 72.4\%) \\\hline
        Other/Prefer not to answer & 27 & 20 & 74.1\% & (57.5\%, 90.6\%) \\ \hline \hline
        Freshman          & 323 & 214 & 66.2\% & (61.1\%, 71.4\%) \\ \hline
        \begin{tabular}[c]{@{}l@{}}Junior or \\ Senior Transfer\end{tabular}                        & 66 & 53 & 80.3\% & (70.7\%, 89.9\%) \\ \hline
    \end{tabular}
    \vspace{-2mm}
    \caption{Demographic distribution of students in our sample. Column (A): total counts across demographic groups, Column (B): counts across software-matched groups, Column (C): percentage of students within subgroup who requested software-assigned study groups, Column (D): population-level confidence intervals on percentages in column (C).}
    
    \label{fig:self_form_comp}
    \vspace{-4mm}
\end{figure}